\documentclass{aastex}
\usepackage{emulateapj5}
\usepackage{onecolfloat5}
\usepackage{apjfonts}
\usepackage{epsfig}

\def\gsim{\;\rlap{\lower 2.5pt
 \hbox{$\sim$}}\raise 1.5pt\hbox{$>$}\;}
\def\lsim{\;\rlap{\lower 2.5pt
   \hbox{$\sim$}}\raise 1.5pt\hbox{$<$}\;}
\newcommand{\be}{\begin{equation}}
\newcommand{\beq}{\begin{equation}}
\newcommand{\ba}{\begin{eqnarray}}
\newcommand{\ee}{\end{equation}}
\newcommand{\eeq}{\end{equation}}
\newcommand{\ea}{\end{eqnarray}}

\newcommand{\msun}{$M_{\odot}$}
\newcommand{\wmap}{{\it WMAP }}

\usepackage{natbib}

\begin{document}
\twocolumn[
\submitted{Submitted to ApJ}
\title{Photoionization Feedback in Low--Mass Galaxies at High Redshift}
\author{Mark Dijkstra, Zolt\'an Haiman}
\affil{Department of Astronomy, Columbia University, 550 West 120th Street, New York, NY 10027, USA}
\vspace{-0.5\baselineskip}
\author{Martin J. Rees}
\affil{Institute of Astronomy, Madingley Road, Cambridge, CB3 0HA, UK}
\vspace{-0.5\baselineskip}
\author{David H. Weinberg}
\affil{Department of Astronomy, The Ohio State University, Columbus, OH 43210, USA}
               
\begin{abstract}
The cosmic ultraviolet (UV) ionizing background impacts the formation
of dwarf galaxies in the low--redshift universe ($z\lsim 3$) by
suppressing gas infall into galactic halos with circular velocities up
to $v_{\rm circ}\sim 75\,{\rm km\,s^{-1}}$.  Using a one--dimensional,
spherically symmetric hydrodynamics code (Thoul \& Weinberg 1995), we
examine the effect of an ionizing background on low--mass galaxies
forming at high redshifts ($z\gsim 10$). We find that the importance
of photoionization feedback is greatly reduced, because (1) at high
redshift, dwarf--galaxy sized objects can self--shield against the
ionizing background, (2) collisional cooling processes at high
redshift are more efficient, (3) the amplitude of the ionizing
background at high redshift is lower, and (4) the ionizing radiation
turns on when the perturbation that will become the dwarf galaxy has
already grown to a substantial overdensity.  We find that because of
these reasons, gas can collect inside halos with circular velocities
as low as $v_{\rm circ}\sim 10\,{\rm km\,s^{-1}}$ at $z>10$.  This
result has important implications for the reionization history of the
universe.
\end{abstract}]


\section{Introduction}
     
The recent detection of a large optical depth to electron scattering
($\tau_e=0.17 \pm 0.04$) by the Wilkinson Microwave Anisotropy Probe
({\it WMAP}, Bennett et al. 2003) suggests that the first sources of
light significantly ionized the intergalactic medium (IGM) as early as
redshift $z\sim 17\pm 4$ (Kogut et al. 2003; Spergel et al. 2003).  A
number of papers (see a recent review and summary in Haiman 2003) have
tried to reconcile this observation with the preceding discovery of a
significant global neutral hydrogen fraction inferred from
Gunn-Peterson troughs in quasars at redshift $\sim 6.3-6.4$ (Becker et
al. 2001; Fan et al. 2003; White, Becker, Fan \& Strauss. 2003).
Haiman \& Holder (2003) and Onken \& Miralda-Escud\'e (2003), in
particular, have emphasized the role that radiative feedback
mechanisms can play in regulating the evolution of the global
emissivity of ionizing radiation, and simultaneously accounting for
these two apparently conflicting observational results.  What is
suggested by these observations is that reionization is completed at
redshifts of $z\approx 6-7$, but that in addition, there is a
significant ``tail'' of (perhaps only partial) ionization extending to
much higher redshift ($z\sim 17$).

If the reionization history is regulated by radiative feedback
effects, such a tail of ionization can naturally be produced.  The
first ionizing sources form in low--mass dark matter (DM) halos with
virial temperatures between $\sim 10^2$ and $10^4$K around $z=20$, in
which cooling by ${\rm H_2}$ molecules can bring the Jeans mass down
to several hundred \msun.  The UV photons produced by these early
population of stars or miniquasars photo-ionize and photo-heat the gas
surrounding them (Shapiro et al. 2003; Oh \& Haiman 2003) and
dissociate ${\rm H_2}$, the main coolant (Haiman, Rees \& Loeb 1997).
This limits the star formation rate and the contribution of these
halos to reionization (Haiman, Abel \& Rees 2000).

The global star--formation can increase again once more massive halos
form, with virial temperatures of $T_{\rm vir}\gsim 10^4$K (or
circular velocities of $v_{\rm circ}\gsim 10\,{\rm km\,s^{-1}}$). In
these halos, gas can condense and cool via an initial phase of atomic
cooling (Oh \& Haiman 2002).  In practice, however, previous models of
the reionization history have excluded star formation in halos up to
much higher virial temperatures, with $T_{\rm vir}\gsim 10^5$K
($v_{\rm circ}\gsim 50\,{\rm km\,s^{-1}}$) in regions of the IGM that
have already been ionized (e.g. Haiman \& Loeb 1997, 1998; Wyithe \&
Loeb 2003; Haiman \& Holder 2003).  The motivation for this exclusion
is that the collapse and cooling of gas is thought to be significantly
suppressed in these photo--heated regions.  The critical halo size up
to which such a suppression is important has been addressed by several
authors (Efstathiou 1992; Thoul \& Weinberg 1996, hereafter TW96;
Navarro \& Steinmetz 1997; Kitayama \& Ikeuchi, hereafter KI00), who
found a critical value of $v_{\rm circ}\sim 50\,{\rm km\,s^{-1}}$.
The exact value of this threshold plays a crucial role in the
reionization history.  This motivates us in the present paper to
revisit the effects of the UV feedback at high redshift.

It is important to realize that existing studies of the UV feedback
mainly focus on low--redshifts ($z\lsim 3$), addressing the
suppression of the formation of dwarf galaxies in the relatively
nearby universe.  Whether or not the same feedback occurs at high
redshift is especially important to clarify in light of the {\it WMAP}
results. Specifically, a strong photoionization feedback can
significantly delay the completion of reionization, provided the
ionizing photon production efficiency is high in these halos (see
Figure 4 in Haiman \& Holder 2003).  In general, one would expect
photoionization feedback to have less of an effect at high redshift,
because cooling times are shorter, the intensity of the ionizing
background is lower, and halos are more compact and can be
self--shielding.  For example, KI00 included radiative transfer
effects in one--dimensional simulations similar to those of TW96. They
found that this lowered the critical velocities by $\sim 5\, {\rm
km\,s^{-1}}$, at low redshift, and that this decrease was more
significant ($\sim 5\, {\rm km\,s^{-1}}$) at high redshifts.  In
three--dimensional simulations, Gnedin (2000) found that the
characteristic mass scale for suppressing the gas fraction in
non--linear halos decreases towards high redshifts, with a dependence
that roughly corresponds to a fixed circular velocity of $20$ km/s at
all redshifts (and provided a fitting formula that has been adopted in
the reionization models of Cen 2003).

In the present paper, we extend previous calculations of the
photoionization feedback to high redshifts, using spherically
symmetric simulations adapted from Thoul \& Weinberg (1995).  Our goal
is to quantify the suppression of gas infall into halos that form at
$z\sim 10$, and to elucidate the physics that distinguishes the
effects of the UV background at $z\sim 10$ from that at $z\sim 2$.
The recent \wmap results suggest that an ionizing background is
already in place at high redshift, and much of early structure
formation takes place under the influence of an ionizing background.
Previous works nevertheless suggests a diminishing importance of the
background towards high redshifts (KI00, Gnedin 2000). In this paper,
we confirm these results, and explicitly compute the change in the
characteristic circular velocity with redshift.  Using a range of
simulation runs, we also quantify the relative importance of different
physical effects that determine the condensed gas fractions we
obtain. Finally, we discuss the significance of (the lack of) feedback
for the high redshift reionization history in light of the \wmap
results.

The rest of this paper is organized as follows. In \S~\ref{sec:code},
we give a brief description of the code we adapted, together with a
few modifications we made for the present application.  In
\S~\ref{sec:result}, we describe the results of our runs, which are
then discussed in more detail in \S~\ref{sec:discuss}. In
\S~\ref{sec:conclude}, we summarize our conclusions and the
implications of this work.  Throughout this paper, we adopt the
background cosmological parameters as measured by the {\it WMAP}
experiment, $\Omega_m=0.27$, $\Omega_{\Lambda}=0.73$,
$\Omega_b=0.044$, $h=0.71$ (Spergel et al. 2003).

\section{Description of the Code}
\label{sec:code}

The code we use is a modified version of a one--dimensional,
spherically symmetric code (Thoul \& Weinberg 1995, hereafter TW95).
It evolves a mixture of dark matter and baryon fluids by moving
concentric spherical shells of fixed mass in the radial direction.  A
description of the numerical method and details of the code can be
found in the above reference.  We start all runs with an initial
density profile following the average shape of a 2$\sigma$ peak in a
Gaussian random field (Bardeen et al. 1986; BBKS).  The initial
density profile is sampled by 6000 dark matter and 1000 gas shells out
to the outermost shell at $r\approx 2.3r_f$, where $r_f$ is the filter
radius for the BBKS profile.  We have explicitly verified that
increasing and decreasing the number of shells by a factor of three
changes our answers for the cold gas fraction below by at most 5
percent. This will introduce an error in the circular velocities of at
most $\sim 3$ km/s.  The total mass (baryons + dark matter) of the
object collapsing at redshift $z=11$ is denoted by $M_f$.  This mass
is initially contained in a sphere of radius $2r_f$, in which we fix
the mean overdensity to be
$\bar{\delta}\equiv(\rho-\bar{\rho})/\bar{\rho}=0.136$. Note that this
places the redshift of collapse at
$z_c\approx(0.136/1.69)\times(1+z_i)-1$, where $z_i$ is the initial
redshift.  We chose $z_i$ based on the desired collapse redshift
(e.g., $z_i=160$ for $z_c=11$).  The total mass of the collapsing
object in our simulation is $1.5 M_f$ (see TW95 and TW96 for more
details).

The code was originally written for a standard cold dark matter (SCDM)
cosmology ($\Omega_m=1,\Omega_{\Lambda}=0$, $H_0=100$ km/s/Mpc).  In
this paper, we adopted the $\Lambda$CDM cosmology favored by the {\it
WMAP} results. However, we are concerned with the collapse of objects
at high redshift, where the dynamical effect of the dark energy is
negligible. In practice, we therefore needed to change only the ratio
$\Omega_{m}/\Omega_b$ (which sets the number ratio of dark matter and
gas shells), and the time-redshift relation ($t=425[(1+z)/12]^{-3/2}$
Myr in the redshift range $2<z<160$).  Because the time--scale for
Compton cooling off the cosmic microwave background (CMB) is shorter
than the Hubble time for $z \gsim 6.5$, we added Compton cooling to
the other cooling terms in the code (e.g., Katz, Weinberg, \&
Hernquist, 1996), \be \Lambda_{\rm Compt}=1.017 \times
10^{-37}(T_e-T_{\rm cmb})T^4_{\rm cmb} n_e \,\,\,\,\, {\rm
erg~sec^{-1}~cm^{-3}}, \ee where $n_e$ is the number density of
electrons, $T_e$ the temperature of the electrons (equal to the gas
temperature), and $T_{\rm cmb}=2.725(1+z)$ is the temperature of the
CMB (Fixsen \& Mather 2002).  Finally, it will be useful to recall the
relation between the circular velocity $v_{\rm circ}$ of a dark matter
halo and its virial temperature:
\begin{equation}
T_{\rm vir}=3.2 \times 10^4\Big{(}\frac{\mu}{0.6}\Big{)}
\Big{(}\frac{v_{\rm circ}}{30}\Big{)}^2\textrm{\hspace{2mm}} \,{\rm K}.
\end{equation}
Below, we study the collapse of galaxies with circular velocities in
the range $10-100$ km/s, for which the mean molecular weight is
$\mu=0.59$.  Note that below $\sim$ 10 km/s, $H_2$ molecule formation
and cooling has to be considered. The effect of the UV background on
such minihalos has recently been studied elsewhere (Oh \& Haiman
2003).  The circular velocity of halos is related to their mass and
collapse redshift (in the absence of any pressure or radiation field)
as 
\be v_{\rm circ}=18.4 \bigg{(} \frac{M_f}{10^{8}M_{\odot}}
\bigg{)}^{1/3} \bigg{(} \frac{1+z_c}{12}\bigg{)}^{1/2}\bigg{(}
\frac{\Omega_m h^2}{0.135}\bigg{)}^{1/6}\textrm{\hspace{2mm} km/s},
\label{eq:vcirc}
\ee 
where $M_f$ is the mass of the halo that collapses at redshift $z_c$.

\section{Runs and Results}
\label{sec:result}

Although the \wmap data suggest that there is an ionizing UV
background at $z \sim 17$, its evolution is poorly constrained.  For
simplicity, in our simulations we adopt a power--law ionizing
background that is described by two parameters,
$J_{\nu}=J_{21}(\nu/\nu_{L})^{-\alpha} \times 10^{-21}$
erg/s/Hz/cm$^2$/sr, where $\nu_L$ is the Lyman limit frequency.  In
addition, the postulated ionizing background is assumed to turn on
suddenly at the location of the collapsing object at a redshift
$z_{\rm on}$.  This sudden turn--on is appropriate before reionization
is complete, i.e. when the Str\"omgren surface from a single ionizing
source passes by a newly forming halo (see more discussion of the
adopted background below). We consider two different amplitudes
($J_{21}=1$ or $J_{21}=10^{-2}$, justified below) and the spectral
slope $\alpha=1$, characteristic of quasars or massive, metal--free
stars. The choice of $\alpha=5$ would be more appropriate for a normal
stellar population (TW96), but since making the spectrum softer would
only reduce the importance of feedback further, we are conservative in
our choice of $\alpha=1$.  For the turn--on redshift, we adopt $z_{\rm
on}=17$ (motivated by \wmap) or $z_{\rm on}=\infty$ (to obtain a
maximal effect).  We run each simulation to redshift $\sim 6$.  In
each simulation run, following TW96, we define the fraction $f_{\rm
coll}=M_c/M_c(p=0)$ of the baryonic mass that has collapsed ($M_c$) at
redshift $z=11$, relative to the amount of baryonic mass that would
have collapsed in absence of pressure
($M_c(p=0)=(\Omega_{b}/\Omega_{m})M_f$).  Since we are considering
halos well above the cosmological Jeans mass, in the absence of a
radiation field we expect $f_{\rm coll}=1$.  A background flux has an
increasingly larger effect on smaller halos, so that in the presence
of a radiation field, we expect $f_{\rm coll}$ to increase
monotonically with circular velocity.

\begin{figure*}[t]
\vbox{ \centerline{\epsfig{file=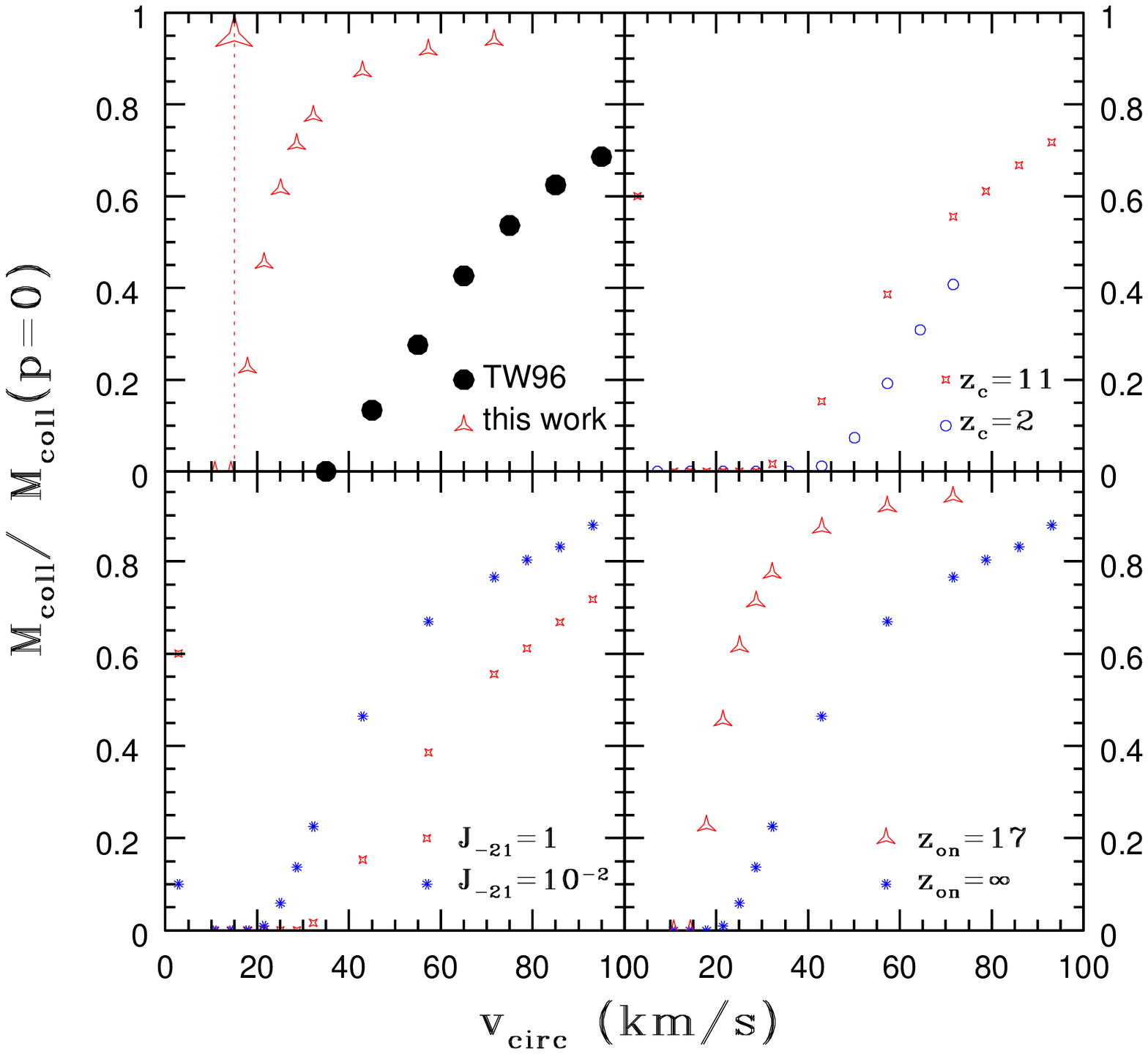,width=18.0truecm}}}
\caption{The fraction $f_{\rm coll}$ of gas that has collapsed as a
function of circular velocity, normalized to the mass $M_c(p=0)$ of
gas that would have collapsed in the same time in the absence of any
pressure.  {\it Upper left panel:} Circles show halos collapsing at
$z=2$, while triangles represent halos that collapse at $z_c=11$.
Feedback becomes much less important at higher redshifts.  This is
even more so when self shielding is considered: The triangle at
$v_{\rm circ}=15$ km/s moves from $f_{coll}=0$ upto $\sim 1$ when
radiative transfer is taken into account.  In the next three panels,
we investigate the reasons for this difference, changing the
parameters of the low--redshift model, one--by--one in succession,
until we arrive back at the high redshift model.  {\it Upper right
panel:} increased cooling rate at higher redshift and densities. {\it
Lower left panel:} lowering the amplitude of the UV-flux; {\it lower
right panel:} late turn--on of the flux (see text for discussion).  }
\label{fig:TW} 
\end{figure*}

In Figure~\ref{fig:TW}, we show the collapsed fraction as a function
of circular velocity under various assumptions.  To begin, we first
reproduced the results of TW96, by adopting their redshift--evolution
for the background flux, and considering halos collapsing at $z_c=2$
in their SCDM cosmology.  Our results for this case are shown by the
the circles in the upper left panel of Figure~\ref{fig:TW}.  The
circles show that gas infall in systems with circular velocities of
$\sim 35$ km/s is suppressed almost completely, while the collapse of
structures with circular velocities of $70$ km/s is suppressed by 50\%
(Note that the often quoted value of $50$ km/s is obtained by defining
$f_{coll}$ at the time $2t_c$.  Here we consider $1t_c$ instead, as
explained in \S~\ref{sec:discuss-clustering}).

Hereafter, we will denote these two velocities by $v_{0}$ and
$v_{1/2}$, respectively.  To facilitate comparison with previous
studies, throughout this paper we focus on $v_{1/2}$, which conveys
the size of a halo that is strongly affected by the radiation.  In the
context of reionization, a question that may be more relevant is
whether {\it any} gas may be available for star formation in the
halo. The answer to the latter question is better conveyed by $v_{0}$,
for which we also quote values below.

The rest of Figure~\ref{fig:TW} describes collapse calculations in the
$\Lambda$CDM cosmology.  First, the triangles in the upper left panel
show the main result of this paper, and correspond to halos that
collapse at $z_c=11$. We use the set of parameters $J_{21}=10^{-2}$,
$\alpha=1$, and $z_{\rm on}=17$.  As the figure clearly shows, the
values of $v_{0}$ and $v_{1/2}$ change dramatically relative to the
low--redshift case: we obtain $v_{0} \approx 15$ km/s, and
$v_{1/2}\approx 20$ km/s.  In the remaining three panels of
Figure~\ref{fig:TW}, we investigate the physical reasons for the
change in the critical circular velocities, by varying each parameter
in succession.  Our approach is to start from a fiducial set of
parameters that describe the typical low--redshift case: $z_c=2$,
$J_{21}=1$, $\alpha=1$, and $z_{\rm on}=\infty$ (in practice, $z_{\rm
on}=z_i$), and then change each parameter in succession, until we
arrive at the high redshift model described by $z_c=11$,
$J_{21}=10^{2}$, and $z_{\rm on}=17$.  The order in which we change
the parameters is arbitrary (and could somewhat modify the relative
importance of each effect).

\noindent{\it Gas Density:} Cooling through two--body processes, such
as recombination and collisional excitation cooling, is more efficient
at high densities.  The influence of increasing the density by a
factor of $(12/3)^3=64$ on the collapsed fraction is shown in the
upper right panel, by adopting $z_c=11$ (stars) vs $z_c=2$ (empty
circles).  The other parameters are held at their fiducial values.  As
the panel reveals, the density increase shifts the entire $f_{\rm
coll}$ curve by $\sim 10$ km/s to the left.  In particular, $v_{1/2}$
is decreased from $\sim 80$ to $\sim 65$ km/s (a factor of $\sim
1.2$).

\noindent{\it Radiation Intensity:} At redshift $z\gsim 11$, the UV
background is expected to be significantly lower than at redshift
$z\sim2$. We therefore next decrease the amplitude of the flux by a
factor of 100 to $J_{21}=10^{-2}$.  This value is motivated below in
\S~\ref{sec:discuss-ionbg}. The lower left panel shows the result of
this decrease (asterisks) vs. $J_{21}=1$ (stars, reproduced from the
upper right panel).  Lowering the flux yields an additional factor of
$\sim 1.4$ decrease in $v_{1/2}$ from $\sim 65$ to $\sim 45$ km/s.

\begin{figure*}
\vbox{ \centerline{\epsfig{file=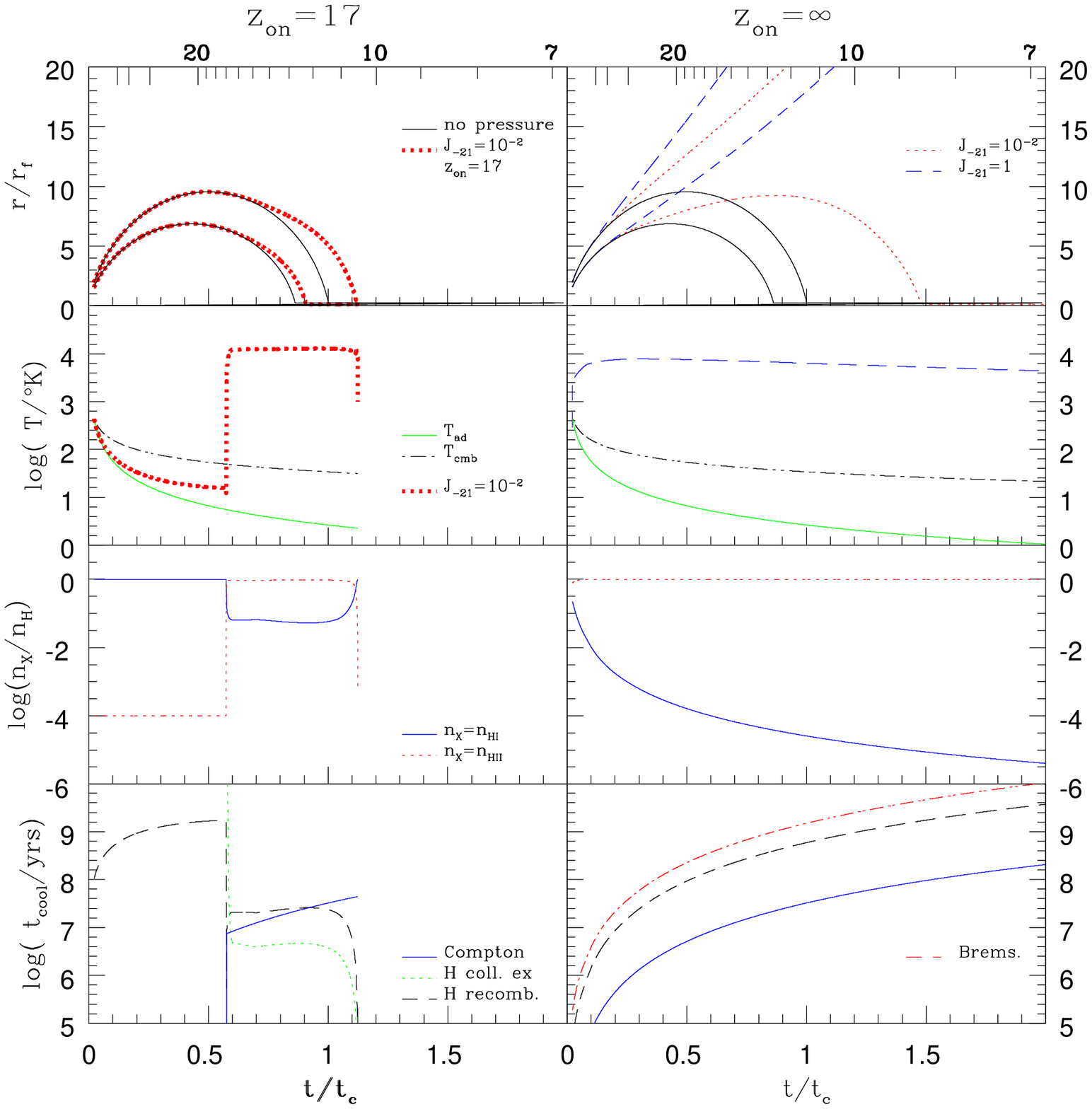,width=18.0truecm}}
\caption{ 
The figure shows (upper to lower panels, respectively) the
evolution of the radii, temperatures, ionization fractions, and
cooling times, of shells during the collapse of a halo with $v_{\rm
circ}=30$ km/s.  The set of panels on the left/right assumes $z_{\rm
on}=17/\infty$.
{\it Top row:} The left panel shows radii of gas mass shells initially
enclosing masses of $0.5M_f$ and $M_f$ (shell 0.5 and 1), with
$J_{21}=10^{-2}$ and in the absence of any pressure (pair of solid
curves).  The thick dotted curves illustrate the effect of a late
turn--on for the flux. In the right panel, the dotted/dashed curves
show the case with $J_{21}=10^{-2}/1$. The solid curves, that show the
pressureless collapse, are reproduced from the left panel.  The panels
below this all focus on shell 1, with the set on the left/right side
showing the $J_{21}=10^{-2}/1$ case.
{\it Second row:} The temperature evolution of the gas shell is shown
together with the CMB temperature (solid curve), and an adiabatic
temperature evolution (dot--dashed curve).
{\it Third row:} The abundances of neutral and ionized hydrogen are
plotted.
{\it Bottom row:} Cooling timescales for the three dominant cooling
processes; Compton cooling (solid curve), recombination (dashed curve)
and excitation cooling (dotted curve). In the left panel, collisional
cooling of neutral hydrogen dominates. Before $z_{\rm on}$ the cooling
time for this process is $\gg 10^{10}$ yrs and rapidly drops to a few
times $10^7$ beyond $z_{\rm on}$.  In the right panel, Compton cooling
dominates. Collisional cooling is very inefficient and is off the
plot.}
\label{fig:panel}}
\vspace*{0.5cm}
\end{figure*}

\noindent{\it Turn-on time of ionizing flux:} A primordial
perturbation that collapses at redshift $z\sim 11$ is already turning
around at $z\sim 18$.  As a result, one expects that such a
perturbation evolves significantly without the presence of any UV flux
before any flux turns--on (see discussion below).  To include this
effect in a simple way, in the lower right panel we change $z_{\rm
on}=\infty$ (empty circles, reproduced from the lower left panel) to
$z_{\rm on}=17$ (triangles; we have arrived back at the ``original''
curve shown in triangles in the upper left panel).  The late turn--on
results in an additional factor of $\sim 2.2$ decrease in $v_{1/2}$
from $\sim 45$ to $\sim 20$ km/s.  Note that photoheating of halos
that collapse at redshift $z\approx 2$ (such as the cases considered
in TW96) starts well before they turn around, and the precise turn--on
redshift is unimportant.  The strong influence of the turn--on time in
our high--redshift collapse case can also be seen directly in the
evolution of shells in the upper pair of panels in
Figure~\ref{fig:panel}.

In Figure~\ref{fig:panel}, we investigate the various physical effects
determining the strength of the UV feedback in more detail. The figure
shows (upper to lower panels, respectively) the evolution of
individual shells, together with their temperatures, ionization
fractions, and cooling times, during the collapse of a halo with
$v_{\rm circ}=32$ km/s.  The set of panels on the left assume that the
UV background is switched on at $z_{\rm on}=17$, and the panels on the
right describe runs in which the flux is on during the entire
simulation.  All quantities are shown as a function of cosmic time
normalized to the collapse time $t_c=425$Myr (the age of the universe
at $z=11$; the upper horizontal axes show corresponding redshifts).
We next describe the individual panels.

\noindent{\it Evolution of gas shells.}  The uppermost row of panels
shows the evolution of two gas shells, initially enclosing masses of
$0.5M_f$ and $M_f$ (we will hereafter refer to these shells as shell
0.5 and shell 1).  The radii are shown in units of the filter radius
$r_f$.  The pair of solid curves shows the evolution of shells in the
absence of any pressure in both panels. Shell 1 collapses
approximately at $t/t_c=1$ as desired. \footnote{The exact collapse
time is sensitive to the initial velocities assigned to the shells. We
adopted velocities derived from linear theory for the growing modes:
$v_i=Hr_i(1-\frac{\bar{\delta_i}}{3})$, where $\bar{\delta_i}$ is the
mean initial overdensity enclosed by shell $i$ (see TW96 for
details).} The dotted and dashed curves in the right panel correspond
to $J_{21}=10^{-2}$ and $J_{21}=1$, respectively, showing that the
shells collapse later (or do not collapse at all) as the the intensity
of the UV background is increased. In the left panel, the thick dotted
line shows the evolution of the shells when the UV background (with
$J_{21}=10^{-2}$) is turned on at redshift $z=17$. At this redshift,
both shell 1 and 0.5 have already turned around. The figure shows that
shell 0.5 was further along its evolution that shell 1, and therefore
the background has a smaller influence in its subsequent evolution
than in that of shell 1.

\noindent{\it Evolution of gas temperature.}  The second row of panels
from the top shows the evolution of the temperature of gas shell 1,
with the dotted curve in the left panel representing the cases with
$J_{21}=10^{-2}$ and the dashed curve on the right when $J_{21}=1$.
For reference, we also show the CMB temperature (dot-dashed curves),
and the adiabatic temperature evolution ($\propto(1+z)^2$, solid
curves).  Initially, heating takes place through adiabatic compression
of the gas shells.  Some heating will take place through inverse
Compton scattering of the CMB photons by electrons, however this is
very inefficient.  We set the fraction of free electrons to
$n_e/(n_{\rm HI}+n_{\rm HII})=10^{-4}$ (the approximately constant
value expected after decoupling, e.g., Peebles 1993).  The left panel
shows that the gas temperature is locked between the adiabatic the CMB
temperature and an adiabatic evolution, until the background turns on.
At $z_{\rm on}$, the shell is rapidly photo--heated to $10^{4}$K,
irrespective of the amplitude of $J_{21}$ (this is enforced by the
assumption of ionization equilibrium in our code; however, this
assumption is justified since the photoionization time--scale is
short, $\sim 10^6$ yr for $J_{21}=10^{-2}$).
 
\noindent{\it Evolution of ionization fractions.}  The third row of
panels from the top shows the evolution of the neutral (solid curves)
and ionized (dashed curves) fractions of hydrogen for shell~1.  The
left panel shows the model with $J_{21}=10^{-2}$ and with the UV
background switched on at $z=17$. As mentioned above, the
free--electron fraction is set to $10^{-4}$ before the UV background
turns on. Once the flux turns on, the gas is highly ionized.  However,
as shown by the solid curve in the left panel, significant neutral
hydrogen remains in this case.  This implies that collisional
excitation cooling can dominate over other cooling processes, and also
that self--shielding effects can be important (see discussion below
and in \S~\ref{sec:discuss-opt}).  For clarity, the right panel only
shows the case with $J_{21}=1$, and demonstrates that the halo gas is
quickly highly ionized, with the neutral fraction declining due to the
dilution of the density in the expanding shell.

\noindent{\it Evolution of cooling timescales.} The dominant cooling
mechanisms are Compton, recombination and excitation cooling.  The
timescales for all processes are shown in the bottom row of panels for
shell 1.  The left panel shows that for shells that do collapse,
collisional excitation cooling (mostly by Lyman $\alpha$ emission)
dominates over recombination and Compton cooling.  This is because the
densities of the shells shown in the left panel are higher (the UV
flux is lower), and the fraction of neutral hydrogen is higher, than
in the corresponding curves in the right panel.  The right panel shows
that for the $J_{21}=1$ case, Compton cooling dominates at high
redshift, followed by recombination. Collisional cooling of neutral
hydrogen is so inefficient that it is off the plot.  The fact that
Compton cooling is dominating over the other cooling mechanisms is
because this shell never collapses and remains at relatively low
densities.  In this case, Bremsstrahlung also contributes to the total
cooling rate at the percent level, as shown by the dot--dashed curve
in the lower right panel.
 
\section{Discussion}
\label{sec:discuss}

The main result of the previous section is that the suppression of gas
infall by photoionization heating is much less severe in halos
collapsing at $z=11$ than at $z=2$. At redshift $z=2$ the presence of
a UV background photo--heats the baryons to $\sim 10^4$ K, which
prevents a fraction of the baryons from condensing inside dark matter
halos. This fraction is $50\%$ for halos with a circular velocity of
$v_{1/2}\approx 75$ km/s, in agreement with previous results by TW96.

We found that for halos collapsing at redshift $z=11$, the critical
velocity is reduced to $v_{1/2}=20$ km/s. This is a change in
$v_{1/2}$ by a factor of $3.5$, corresponding to a decrease in the
halo mass by a factor of $3.5^3\approx 42$ (eq.~\ref{eq:vcirc}).  We
found that several factors contribute to this reduction.  The largest
factor of $\approx 2.2$ came from our assumption that for halos
collapsing at $z=11$, the UV background is not always ``on'' - we
choose to turn on the flux when the perturbation has just turned
around (see the lower right panel in Figure~\ref{fig:TW}).  This
corresponds to $z_{\rm on}=17$, the redshift where {\it WMAP}
indicates that a significant fraction of the volume of the IGM is
ionized.  At high redshift, the densities are higher, and therefore
cooling is more efficient. This was responsible for a factor of
$\approx 1.2$ (see the upper right panel in Figure~\ref{fig:TW})). A
comparable factor ($\approx 1.4$) came from lowering the flux $J_{21}$
from $1$ to $10^{-2}$ (in the next section, we discuss our choice for
the amplitude $J_{21}=10^{-2}$).  In comparison, TW96 found that the
influence of photoionization came from the suppression of turnaround
and collapse by heating, rather than from reduction in cooling rates,
and that therefore the amplitude of the background was unimportant
(though the slope was significant because of its influence on
temperature).  This agrees well with our conclusion that the late
turn--on of the background flux (relative to $t_c$), accounts for most
of the reduction in $v_{1/2}$.  However, in our high redshift case,
the effect of photoionization on cooling does have an additional
impact, and the amplitude of the flux is therefore important, as well.

Two of the simplifying assumptions in our treatment are that the gas
is optically thin and in ionization equilibrium.  In reality, when the
collapsing gas is illuminated by a flux that turns on suddenly, the
initial heating rate can significantly exceed the values we assume.
This is because neither the extra heat input from the initial
ionization of each atom (e.g., Hui \& Gnedin 1997), nor the temporary
hardening of the spectrum due to absorption by the large initial
column of neutral atoms (Abel \& Haehnelt 1999), are captured by our
treatment.  However, efficient cooling at high redshift tends to
mitigate these effects. In addition, we note that the gas in our
high--redshift collapse runs, is never highly ionized, and so the
equilibrium heating rates are not significant underestimates.  In
order to explicitly verify these assertions, we performed a run
analogous to the fiducial high--redshift run shown in the left panels
in Figure~\ref{fig:panel}, except that (1) we artificially set the
temperature at $z_{\rm on}$ to 50,000 K (the highest temperature to
which the gas could get photo--heated due to self--shielding effects),
and (2) then follow the non--equilibrium evolution of the ionization
fractions and heating rates.  We find that these effects decrease
$f_{\rm coll}$ by less than $1\%$.

\begin{figure*}
\vbox{ \centerline{\epsfig{file=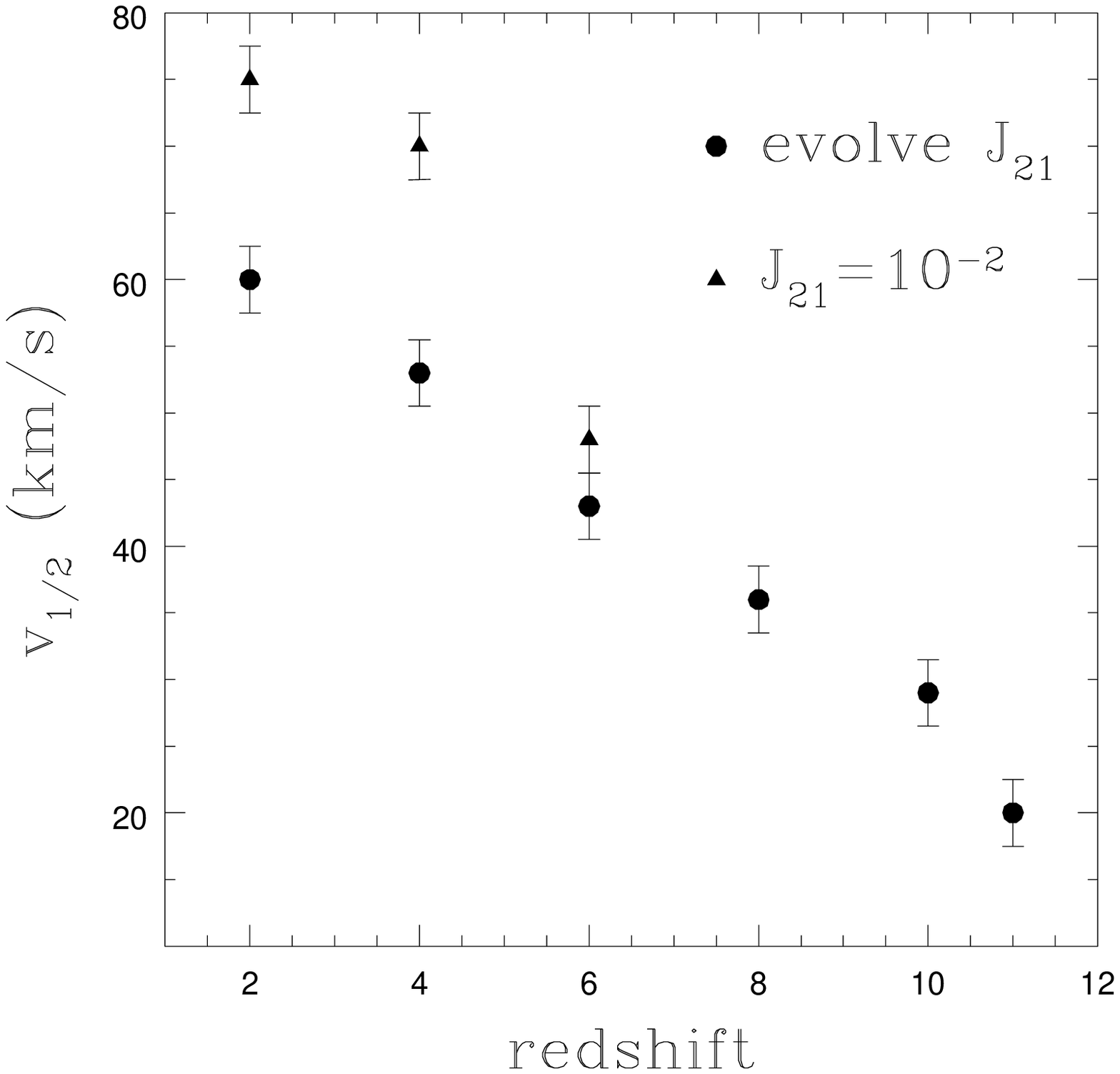,width=8.0truecm}}
\caption{The evolution of $v_{1/2}$ as a function of redshift. Circles
denote the case with $J_{21}=10^{-2}$ for all redshifts greater than
$z_{\rm on}$. Triangles denote the case when $J_{21}$ evolves as a
function of redshift. }
\label{fig:vvsz}}
\vspace*{0.5cm}
\end{figure*}

So far our main focus was to differentiate the value of $v_{1/2}$ at
$z=2$ and $z=11$. In semi-analytic models, it is useful to know how
$v_{1/2}$ changes as a function of redshift. This is shown in
Figure~\ref{fig:vvsz}.  The circles denote the case with
$J_{21}=10^{-2}$ for all redshifts (the error--bars show the range
obtained using a different number of shells $N_{\rm DM}=2000, 6000,
18000$).  Since this flux is not appropriate at lower redshifts, the
triangles denote the case with the UV background evolving as a
function of redshift The evolution for $J_{21}$ is taken from KI00
(their eq.~4), and has $J_{21}=1$ for $3 \leq z \leq 6$ and
monotonically decreasing after this to $\sim 10^{-2}$ at $z \sim 14$.
This demonstrates that at lower redshifts $z \leq 6$ the value of
$v_{1/2}$ moves up further, because of the higher value of
$J_{21}$. Since in this evolution scenario, $J_{21}$ is too high at
redshifts $5 \leq z \leq 6$, the triangles may be viewed as upper
limits.  In \S~\ref{sec:discuss-opt} we will discuss that the the
value of $v_{1/2}$ will decrease, when self--shielding and radiative
transfer are included.

\subsection{Ionizing Background}
\label{sec:discuss-ionbg}

As mentioned above, the evolution of the UV background is poorly
constrained beyond $z=6$.  Under the assumption that the IGM is in
photoionization equilibrium with a uniform background flux, several
authors have used the presence of the Gunn--Peterson trough in a
redshift $z=6.28$ quasar (Becker et al. 2001) to place the upper limit
$J_{21}\lsim 0.03$ on the background flux at $z\sim 6$ (McDonald \&
Miralda-Escud\'e 2001; Cen \& McDonald 2002; Fan et al. 2002; Lidz et
al. 2002).  The background at higher redshifts would be expected to be
lower; in general, under the assumption of photoionization equilibrium
implies the relation
\begin{eqnarray}
J_{21}=10^{-2} \Big{(}\frac{2.2\times10^{-3}}{1-x}\Big{)}\Big{(}\frac{3+\alpha}{4}\Big{)}
\Big{(}\frac{\Omega_bh^2}{0.022}\Big{)}\Big{(}\frac{1+z}{18}\Big{)}^3
\label{eq:j}
\end{eqnarray} 

\vspace{0.5\baselineskip}\noindent between the neutral fraction (by
number) of hydrogen atoms $(1-x)$, and the background flux $J_{21}$.
While $1-x$ is believed to be $\gsim 10^{-4}$ at $z\approx 6$, our
assumption of $J_{21}=10^{-2}$ at $z=17$ is equivalent to
$(1-x)=2.2\times10^{-3}$ at this redshift.

A uniform background at $z\sim 17$, however, is likely to be an overly
simplistic assumption. Indeed, our main motivation here for examining
the photoionization feedback effect is to quantify the ability of
halos with $10~{\rm km\,s^{-1}}\lsim v_{\rm circ}\lsim 75~{\rm
km\,s^{-1}}$ to contribute to reionization -- a question that would be
less relevant once the discrete ionized regions percolate to establish
a uniform background, implying that the IGM has already been fully
ionized.  As a result, a picture that is more relevant to consider
(and which is also more realistic), is that of a swiss--cheese
topology for the IGM: discrete HII regions (or perhaps the overlapping
cluster of a few HII regions) populate the otherwise still neutral
IGM.  In this case, we can estimate the UV flux seen by the gas inside
an individual HII region.  For example, let us consider a $10^8 {\rm
M_{\odot}}$ dark matter halo, in which a fraction $f_{*}=0.1$ of the
baryons turns into stars, generating $n_{\rm phot}=4000$ ionizing
photons per stellar baryon at a steady rate over a period of
$t_{sys}=10^8$ yrs, of which a fraction $f_{esc}=0.1$ escapes into the
IGM (these fiducial numbers were adopted from Haiman \& Holder 2003).
At a distance $r$ away from the central source, we then have the flux
\begin{eqnarray}
J_{21}(r)=3.6 \times 10^{-3}\Big{(}\frac{17.9 \textrm{\hspace{1mm} kpc}}{r}\Big{)}^2\Big{(}\frac{f_{*}}{0.1}\Big{)}
\Big{(}\frac{f_{esc}}{0.1}\Big{)}\\ \nonumber
\Big{(}\frac{\Omega_b/\Omega_m}{0.16}  \Big{)} \Big{(}\frac{M}{10^8M_{\odot}}\Big{)}\Big{(}\frac{n_{phot}}{4000} \Big{)}\Big{(}\frac{10^8 yr}{t_{sys}} \Big{)}, \\ \nonumber
\label{eq:j(r)}
\end{eqnarray}
where the fiducial radius $r=17.9$ kpc is the maximum physical radius
the HII region can obtain (in the absence of recombination). More
generally, the maximum physical radius the halo of mass $M$ can ionize
is
\be
R_{HII}=17.9\Big{(}\frac{0.022}
{\Omega_bh^2} \frac{n_{phot}}{4000}\frac{f_{*}}{0.1}\frac{f_{esc}}{0.1}\frac{M}{10^8M_{\odot}}\Big{)}^{1/3}\Big{(}\frac{18}{1+z}\Big{)}
\textrm{\hspace{2mm} kpc}.
\label{eq:HII}
\ee 
It is most likely to locate another halo in the edge of an HII sphere,
simply because the volume is largest there.  We conclude that the
ionizing flux seen by a collapsing object at $z\gsim 10$ is small,
$J_{21} \lsim 10^{-2}$, regardless of whether or not the individual
HII regions have overlapped.

It is also interesting to consider the effect of spatial correlations
between ionizing sources and forming halos.  The mean separation
between halos of mass $M=10^8~{\rm M_\odot}$ at $z=17$ is $\sim 60$
physical kpc (Jenkins et al. 2001).  Thus, the random chance of
finding a second, similar--size halo within the HII sphere is $\sim
3\%$.  However, the halos are highly correlated (with a bias parameter
$b\sim 7$ (Mo \& White 1996), and a correlation function $\xi(r=20{\rm
kpc})\sim 5$), bringing the probability of finding a second halo
within the HII region to $\sim 10\%$. Furthermore, we are interested
in {\it would--be} halos, in addition to fully formed ones.  The mean
separation between halos which are at their turn--around stage at
$z=17$ is $\sim 3$ times smaller than that of fully formed halos
($\sim 1.5\sigma$ peaks, rather than $\sim 3\sigma$ peaks).  One or
two such halos in their formative stages should therefore be found in
a typical HII region, experiencing photo--ionization feedback.  This
implies, in turn, that the fraction of halos potentially subject to
the feedback is significantly larger than the simple ``volume filling
factor'' of the HII regions.

\subsection{Optical Depth Effects}
\label{sec:discuss-opt}

Another important issue is that towards higher redshifts, halos with a
fixed circular velocity can self--shield against the ionizing
radiation more effectively.  If the gas is in hydrostatic equilibrium
within an NFW halo (Makino, Sasaki, \& Suto (1998), with concentration
parameter $c=5$), a line of sight across a halo with circular velocity
$v_{\rm circ}$ at redshift $z$ has a total hydrogen column density
(from the center to the virial radius of the halo)

\begin{eqnarray}
\nonumber
N_{\rm H}= 2.5 \times 10^{21}
\left(\frac{v_{\rm circ}}{15\,{\rm km\,s^{-1}}}\right)
\left(\frac{1+z}{12}\right)^{3/2}\\ 
\left(\frac{\Omega_bh^2}{0.022}\right)
\left(\frac{0.135}{\Omega_mh^2}\right)^{1/2}\,\,\,\,\,\,{\rm cm^{-2}}.
\label{eq:hcolumn0}
\end{eqnarray} 

\begin{figure*}
\vbox{ \centerline{\epsfig{file=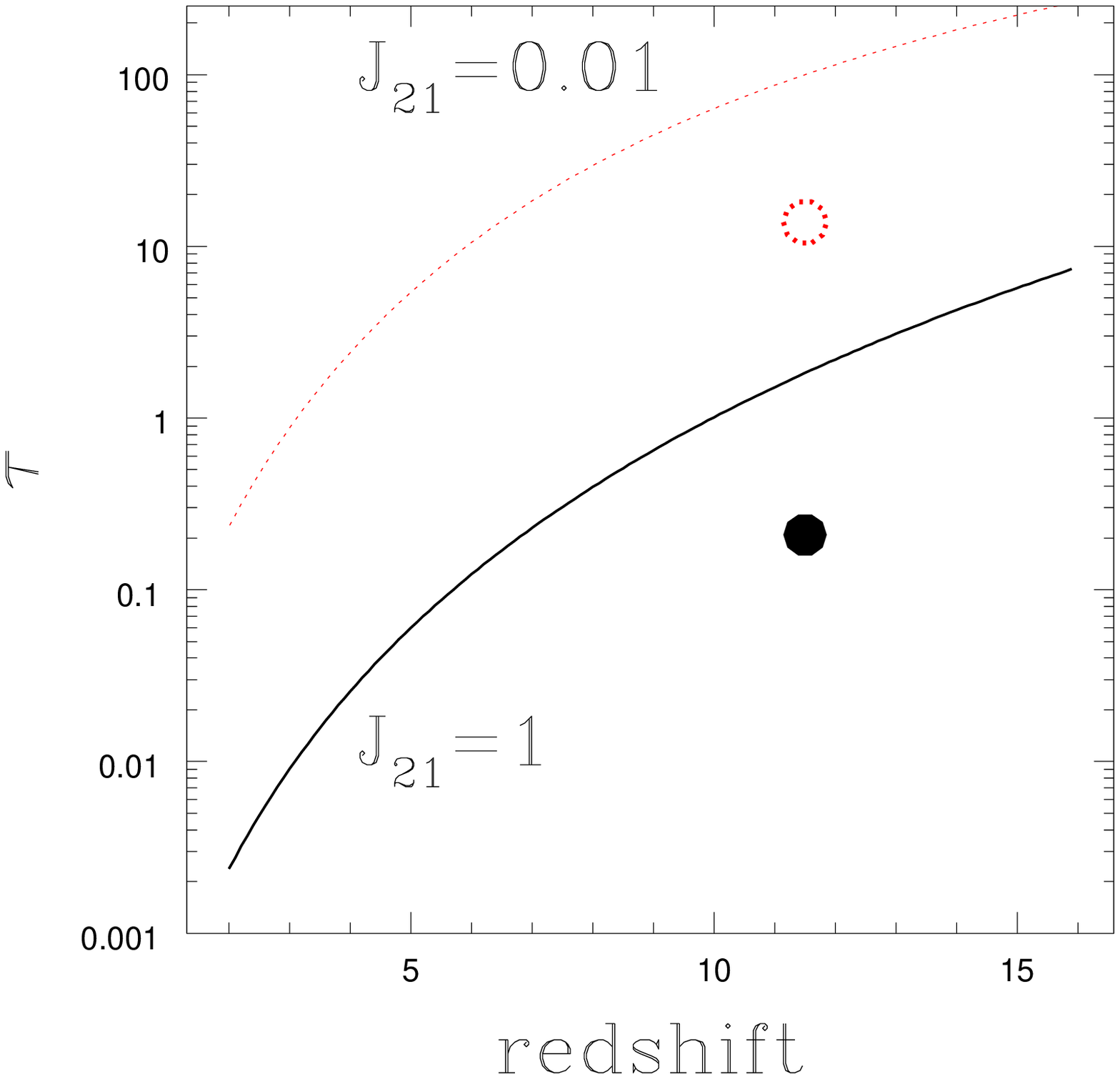,width=8.0truecm}}
\caption{The optical depth from the virial radius inward to half this
radius for 15 km/s NFW DM halos with their gas in hydrostatic
equilibrium. The two curves denote the optical depth when the gas is
in ionization equilibrium with an ionizing background
$J_{\nu}=J_{21}(\frac{\nu_L}{\nu})\times 10^{-21}$ erg/s/cm$^2$/sr/Hz.
Two different values of $J_{21}$ in the figure are shown, as labeled.
The open dashed circles show the optical depth as extracted from the
fiducial high redshift run. The optical depth is evaluated at the
instant the ionizing background is switched on.}
\label{fig:tau}}
\end{figure*}

When such a halo resides in a photo--ionizing background, the gas is
ionized and the column density of neutral atoms is reduced.  In
general, a simple expression analogous to equation~(\ref{eq:hcolumn0})
does not exist in this case.  Here we assume that the gas is optically
thin ($\tau=0$), and in ionization equilibrium with a background flux
with amplitude $J_{21}$. We then show the numerically computed optical
depth ($\tau=N_{\rm HI}\sigma_{\rm HI}$, where $\sigma_{\rm HI}$ is
the photoionization cross section for hydrogen) for $15$ km/s halos as
a function of redshift in Figure~\ref{fig:tau}.  The solid/dotted
curve corresponds to $J_{21}=1/10^{-2}$ (both at the temperature of
$10^4$K).  The curves show the optical depth from shell 1 inward to
shell 0.5, i.e. to the shell containing 50\% of the gas mass, and are
therefore relevant for computing the value of $v_{1/2}$.  The figure
demonstrates that self--shielding can become important at high
redshift: typical column densities at $z\gsim 12$ are several order of
magnitudes higher than at $z\sim 2$, both because of the increased
densities and the lower amplitude of the ionizing background.

Of course, NFW halos are collapsed structures, and therefore have
densities $\gtrsim 178$ times the mean density of the universe at any
given redshift. Therefore they are more efficient self shielders than
the objects in our simulations, which are exposed to a UV background
at turnaround, when the mean density interior to shell 1 is only $\sim
5$ times the mean density.  Furthermore, the curves in
Figure~\ref{fig:tau} implicitly assume that within the virial radius
of the dark matter halo, the mass fraction of baryons is equal to the
universal value, i.e., that a fraction $f_{\rm coll}=1$ of the baryons
has collapsed in the dark matter halo.  In our simulations runs, this
fraction is reduced by the UV background to be less than unity, which
further reduces the optical depth relative to the simple analytical
cases shown in Figure~\ref{fig:tau}.

We therefore next measure optical depth directly in our simulations runs.
Since these runs do not include radiative transfer and assume the gas is
optically thin, we are effectively checking whether these runs are
self--consistent.  We measure the optical depth in runs with $z_{\rm on}=17$,
immediately after the UV background is turned on, and when the gas shells are
just turning around. We find that high--redshift halos can easily reach $N_{\rm
HI}\gsim 10^{18}{\rm cm^{-2}}$ and can therefore self--shield even at the time
of turn--around.  More precisely, in our fiducial high--redshift runs with
$z_c=11$, $J_{21}=10^{-2}$, $\alpha=1$ and $z_{\rm on}=17$, we find that at the
redshift $z_{\rm on}=17$, the smallest halo we simulated ($v_{\rm
circ}=15\,{\rm km\,s^{-1}}$), has a column density of $N_{\rm HI} \sim 2 \times
10^{18}{\rm cm^{-2}}$ between shell 1 (which roughly represents the edge of our
object) and shell 0.5. The corresponding optical depth ($\tau =14$) is shown as
an open dashed circle in Figure~\ref{fig:tau}. Likewise, the lower, solid
circle in Figure~\ref{fig:tau} shows the optical depth at turnaround when
$J_{21}$ is changed to 1.  As expected, the optical depth in both cases is
lower than that for the gas in the NFW halo.
\footnote{When the optical depth to the center of the halos is
calculated, a large contribution comes from the innermost shells,
where the density, and therefore the neutral fraction, is high.
However, the evolution of the innermost few shells is known to depend
on the choice of the numerical boundary condition at $r=0$ (e.g.,
Haiman, Thoul \& Loeb 1996; Forcada-Miro \& White 1997), and we
therefore chose to remain conservative, and to exclude these shells
from our calculation of the optical depth.}  As a result of the high
optical depth ($\tau=14$) in the 15 km/s fiducial high--redshift run,
shell 0.5 will see only a fraction $\exp(-\sigma_{\rm HI}N_{\rm
HI})\sim 10^{-6}$ of the UV background.  This fraction is decreased by
an additional factor of $\sim 10$ when the {\it average} transmission,
$<e^{-\tau}> \equiv \frac{1}{4\pi}\int d\Omega
e^{-\tau(\theta,\phi)}$, is considered.  The above makes it clear that
self--shielding is important at high redshift, and the triangles shown
in the upper left panel of Figure~\ref{fig:TW} should therefore be
regarded only as lower limits.

To demonstrate the importance of shelf shielding more accurately, we
performed one run which takes into account radiative transfer.  We
follow the approach of Kepner, Babul, \& Spergel(1997). To keep the
computing time relatively short, only the opacity due to hydrogen is
included in a self consistent way (ignoring helium). For each shell,
we calculate the transmission, averaged over solid angle, and obtain
the processed spectrum and reduced photo-ionization and heating
rates. Although this approach still underestimates the total opacity,
it gives us much more stringent constraints on $v_{1/2}$ and $v_{0}$.

We show in Figure~\ref{fig:TW} how the point with $v_{\rm circ}=15$
km/s shifts upwards from $f_{\rm coll}=0$ to 1, when the opacity due
to hydrogen is taken into account in the simulations.  Including
radiative transfer further increases the optical depths of our halos,
since the inner shells see (much) lower fluxes than we are assuming in
our optically thin runs, and therefore collapse more rapidly and to
higher densities.  These conclusions are consistent with those of
KI00, who apply solutions of the radiative transfer equation, obtained
in a plane parallel geometry, to spherical 1-D simulations.  This
causes them to overestimate the true optical depth.  They found that
the value of $v_{1/2}$ is decreased by 5 km/s at low redshift, and by
as much as 10 km/s at higher redshift, as compared to the transparent
case studied by us and by TW96.

Finally, as argued above, the question of photoionization feedback
studied in this paper is most relevant to halos forming inside
existing HII regions of a single ionizing source.  The UV radiation
field is then very asymmetric. This is not, of course, captured in our
spherically symmetric simulations. If, however, only one side of the
collapsing halo is exposed to the UV radiation, this will make it
easier for gas on the shadowed side to collapse.  In summary, the
issues of radiative transfer and three--dimensional topology are
likely very important and should be addressed in future work.  Here we
simply note that our main conclusion -- namely that feedback is less
important towards higher redshifts -- is likely to be significantly
strengthened by both effects.

\begin{table}[ht]\small
\caption{Critical circular velocities (in km/s) for gas cooling
defined at two epochs: $t_c$ vs. $2t_c$ and in different runs
corresponding to the different panels in Figure~\ref{fig:TW} (see
\S~\ref{sec:result} for discussion).}
\label{table}
\begin{center}
\begin{tabular}{ccccc}
\tablewidth{3in}
 & $1t_c$: & &$2t_c$: & \\
\hline
\hline
change                    & $v_{1/2}$ & $v_{0}$   & $v_{1/2}$ & $v_{0}$ \\
\hline
$z=2$ data                & 80        & 45        &  55       & 30      \\
raising z (or $\rho$)     & 65        & 30        &  37       & 20      \\
lowering $J_{21}$         & 40        & 20        &  25       & 15      \\
turn on at $z=17$         & 20        & 15        &  20       & 15      \\
self--shielding           & 10        & 10        &  10       & 10      \\
\hline
\end{tabular}\\[12pt]
\end{center}
\end{table}

\subsection{Clustering and Merging}
\label{sec:discuss-clustering}

Another issue is the epoch at which we define to look at how much gas
has cooled and collapsed in a halo.  An often quoted value from TW96
for the critical velocity for suppressing gas infall, $75$ km/s,
refers to $v_{1/2}$ defined at twice the pressureless collapse time,
$2t_c$.  This particular choice can be motivated by the time--scale of
mergers between halos. Typical halos merge into more massive halos on
a timescale that is approximately the Hubble time (Lacey \& Cole
1993), so that requiring gas to cool and collapse by $2t_c$ is roughly
equivalent to requiring this to occur before the halo undergoes
significant mergers (at which stage, gas can, in any case, cool and
condense more effectively into the deeper potential well of the the
newly formed merger product).  In the context of reionization, we are
interested in potentially more prompt feedback, since the global
neutral fraction at high redshift, when the rare, high--sigma peaks
are collapsing, can be evolving on a shorter time--scale ($\lsim 10\%$
of the Hubble time, e.g., Haiman \& Loeb 1997, 1998). It is therefore
useful to know the amount of the baryons available immediately after
the collapse of the halo for star--formation, and thus for
contributing to reionization.  For completeness, the values of
$v_{1/2}$ and $v_{0}$ after $t=t_c$ are shown in Table~\ref{table},
and compared to these values measured after $2t_c$. It can be seen
from this table that the final results for the velocities (bottom row)
remain unchanged. However, the relative importance of the three causes
(\S~\ref{sec:result}) determining their value, and driving them to be
smaller than in our $z=2$ runs, are different (cooling contributing
more when $t_c$ is used).

\subsection{The Impact on the Global Reionization History}
\label{sec:discuss-reionization}

The presence or absence of a strong UV feedback on halos with circular
velocities just above the threshold $v_{\rm circ}\gsim 10\,{\rm
km\,s^{-1}}$ has a strong influence on the global reionization
history.  As argued by Oh \& Haiman (2003; see also Ricotti et
al. 2002; Haiman, Abel \& Rees 2000; Ciardi et al. 2000), halos with
$v_{\rm circ}\lsim 10\,{\rm km\,s^{-1}}$ are unlikely to significantly
contribute to reionization.  As a result, the onset of the
reionization epoch, when a significant fraction of the IGM can first
be ionized, hinges on the efficiency of ionizing photon production in
halos with $v_{\rm circ}\gsim 10\,{\rm km\,s^{-1}}$.  Realistically,
however, the UV feedback cannot have a global impact until a
significant fraction of the IGM is already ionized (although see notes
about clustering above).  On the other hand, the UV feedback --if
important -- would impede the late stages of reionization and
percolation, and delay the epoch when the Gunn--Peterson
'breakthrough' occurs.

To illustrate this more clearly, we can make the following simple
comparison.  Let us assume that halos with circular velocities up to
$v_{\rm circ}=75\,{\rm km\,s^{-1}}$ are excluded from by strong
feedback from completing the reionization epoch (e.g., Haiman \&
Holder 2003; Wyithe \& Loeb 2003).  In this case, percolation will be
achieved by halos with $v_{\rm circ}\geq 75\,{\rm km\,s^{-1}}$.  For
example, using the mass function in Jenkins et al. (2001), we find
that the universal fraction of baryons collapsed into halos of this
size at $z=7$ is $\sim 1\%$. With reasonable choices of the efficiency
parameters (e.g. Haiman \& Holder 2003; Cen 2003), the IGM will be
fully ionized close to this redshift at $6\lsim z\lsim 7$.  In
comparison, if we assume that all halos with $v_{\rm circ}\geq
10\,{\rm km\,s^{-1}}$ can contribute to reionization (representing the
other extreme case of no UV feedback at all), then we find the
collapsed fraction at $z=7$ is $\sim 10\%$, and the fraction of $\sim
1\%$ is achieved already by $z=12$, making percolation feasible at
this earlier epoch. In other words, under the assumption of fixed
efficiencies, the presence/absence of the feedback would make a factor
of $\sim 1.7$ difference for the redshift of the ``break--through''
epoch.  Assuming the IGM is suddenly fully ionized at this
break--through, this translates to a factor of $\sim 2.2$ increase in
the electron scattering optical depth, from $\tau\sim 0.05$ at $z=7$
to $\tau\sim 0.11$ at $z=12$.

We found in this paper that UV feedback is unimportant at high
redshift.  Naively, this would suggest that relatively smaller halos
can contribute to reionization, and help explain the large optical
depth measured by {\it WMAP}.  On the other hand, as explained in the
introduction, the {\it WMAP} results suggest that the reionization
history is more complex -- with a ``percolation'' being completed at
redshifts of $z\approx 6-7$, but with a significant tail of (perhaps
only partial) ionization extending to much higher redshift.  As clear
from the preceding paragraph, a strong UV feedback would, in fact,
have naturally helped to produce this behavior (as argued, e.g., in
Haiman \& Holder -- see Figure 4 in that paper). Indeed, provided
halos near the $\sim 10$ km/s threshold otherwise have a high
efficiency of ionizing photon production, a strong UV feedback would
result in an extended episode during which the IGM is kept partially
ionized. Full percolation could then be achieved only around $z\sim 7$
when halos with $v_{\rm circ}\sim 75\,{\rm km\,s^{-1}}$, impervious to
feedback, start appearing.  The absence of the UV feedback means that
some other physics must then be responsible for the effective decrease
in the global emissivity of ionizing radiation. Such a decrease is
required to explain the complex reionization history suggested by the
combination of {\it WMAP} and lower redshift data.

\section{Conclusions}
\label{sec:conclude}

In this paper, we investigated the importance of photoionization
heating in suppressing gas cooling and infall in halos with circular
velocities in the range $10-100$ km/s, collapsing at high redshift.
We used a modified version of the one--dimensional, spherically
symmetric, Lagrangian hydro code written by Thoul \& Weinberg (1995),
thereby directly extending the earlier calculations by Thoul \&
Weinberg (1996) to redshifts $z>10$.  We reproduced the results of
TW96, namely that UV background has a strong impact in the
low--redshift universe ($z\lsim 3$), suppressing gas infall by a
factor of $\sim 2$ in galactic halos with circular velocities of
$v_{1/2}\sim 50\,{\rm km\,s^{-1}}$.  The main new result of the
present paper is that suppression is much less significant at high
redshift.  For halos collapsing at $z=11$, we found $v_{1/2}=20$ km/s,
and that a few percent of the gas can cool and condense in halos with
circular velocities as low as $v_{0}$=10 km/s.

In \S~\ref{sec:discuss}, we identified three reasons for this result,
namely, (1) the amplitude of the ionizing background at high redshift
is lower, (2) the ionizing radiation turns on when the perturbation
that will become the dwarf galaxy has already grown to a substantial
overdensity, and (3) collisional cooling processes at high redshift
are more efficient.  We find that the relative importance of these
three effects depends on the details of how and at what epoch the
collapsed gas is identified, but we find that the combined effect is to
essentially eliminate strong UV feedback in all halos with $v_{\rm
circ}\sim 10\,{\rm km\,s^{-1}}$.

In our simulations, the halos were assumed to be optically thin, and
spherically symmetric.  We explicitly showed that the halos are, in
fact, self--shielding, and the radiation field is expected to be very
anisotropic (the halos being illuminated by a single nearby source).
Issues of radiative transfer and three--dimensional topology are
therefore crucial in more detailed modeling of the UV feedback studied
here. However, both effects will significantly strengthen our main
result, namely that feedback is largely eliminated in high--redshift
halos.

The presence or absence of a strong photoheating feedback, which
determines whether stars can or cannot form in halos just above the
threshold $v_{\rm circ}\gsim 10\,{\rm km\,s^{-1}}$, has a strong
influence on this reionization history.  While the lack of a UV
feedback, as suggested here, can help explaining the high electron
scattering optical depth detected by {\it WMAP}, it leaves the puzzle
that some other physics must be responsible for the complex
reionization history suggested by combining the {\it WMAP} results
with data from lower redshifts.

\acknowledgements{We thank Jordi Miralda-Escud\'e for useful comments
on the manuscript, and Anne Thoul for the use of the TW96 code. ZH
acknowledges financial support from NSF grant AST-03-07291.}

\end{document}